# Mixing Interstellar Clouds Surrounding the Sun

Paweł Swaczyna[1†], Nathan A. Schwadron[1,2], Eberhard Möbius[2], Maciej Bzowski[3], Priscilla C. Frisch[4], Jeffrey L. Linsky[5], David J. McComas[1], Fatemeh Rahmanifard[2], Seth Redfield[6], Réka M. Winslow[2], Brian E. Wood[7], Gary P. Zank[8]

[1]Department of Astrophysical Sciences, Princeton University, Princeton, NJ 08544, USA.
[2]Space Science Center, University of New Hampshire, Durham, NH 03824, USA.
[3]Space Research Centre PAS (CBK PAN), 00-716 Warsaw, Poland.
[4]Department of Astronomy and Astrophysics, University of Chicago, Chicago, IL 60637, USA.
[5]JILA, University of Colorado and NIST, Boulder, CO 80309, USA.
[6]Department of Astronomy and Van Vleck Observatory, Wesleyan University, Middletown, CT 06459, USA.
[7]Space Science Division, Naval Research Laboratory, Washington, DC 20375, USA.
[8]Department of Space Science, University of Alabama in Huntsville, AL 35805, USA.
[†]Corresponding author: swaczyna@princeton.edu

**Abstract**

On its journey through the Galaxy, the Sun passes through diverse regions of the interstellar medium. High-resolution spectroscopic measurements of interstellar absorption lines in spectra of nearby stars show absorption components from more than a dozen warm partially ionized clouds within 15 pc of the Sun. The two nearest clouds – the Local Interstellar Cloud (LIC) and Galactic (G) cloud – move toward each other. Their bulk heliocentric velocities can be compared with the interstellar neutral helium flow velocity obtained from space-based experiments. We combine recent results from Ulysses, IBEX, and STEREO observations to find a more accurate estimate of the velocity and temperature of the very local interstellar medium. We find that, contrary to the widespread viewpoint that the Sun resides inside the LIC, the locally observed velocity of the interstellar neutral helium is consistent with a linear combination of the velocities of the LIC and G cloud, but not with either of these two velocities. This finding shows that the Sun travels through a mixed-cloud interstellar medium composed of material from both these clouds. Interactions between these clouds explain the substantially higher density of the interstellar hydrogen near the Sun and toward stars located within the interaction region of these two clouds. The observed asymmetry of the interstellar helium distribution function also supports this interaction. The structure and equilibrium in this region require further studies using in situ and telescopic observations.

**Keywords:** Interstellar atomic gas (833), Interstellar clouds (834), Interstellar medium (847), Heliosphere (711), Interstellar absorption (831), Intercloud medium (803), Interstellar medium wind (848), Interstellar dynamics (839)

# 1. Introduction

The local interstellar medium consists of a low-density fully ionized cavity called the Local Bubble extending over 100 pc from the Sun (Frisch & York 1983; Breitschwerdt 1998) and smaller warm (3000 – 13,000 K) partially ionized interstellar clouds inside this cavity (Redfield & Linsky 2008). Local clouds appear to originate from an evolved superbubble associated with Loop I (Frisch et al. 2011). Individual clouds show velocity deviations from a rigid-body flow and indicate that the flow is decelerating as it moves outward in the superbubble (Frisch et al. 2002). The velocity structure of these clouds suggested the presence of nearby interstellar shocks (Grzedzielski & Lallement 1996), possibly aligned with the interstellar magnetic field shaping the heliosphere (Frisch et al. 2022).

The absorption lines toward nearby stars initially revealed two clouds in the direct neighborhood of the Sun called Galactic (G) and Anti-Galactic (AG) clouds based on their position in the sky relative to the Galactic center (Lallement & Bertin 1992). Further observations indicated apparent consistency of the AG cloud flow with the very local interstellar medium (VLISM) observed in the heliosphere, and consequently, the cloud was renamed the Local Interstellar Cloud (LIC) (Bertin et al. 1993; Lallement et al. 1995). Further observations confirmed that the absorption lines toward the stars located close to the Galactic center do not show absorption from the LIC, implying that the Sun must be within a small fraction of a parsec from the LIC edge (Wood et al. 2000). Nevertheless, the equivalence between the LIC and the VLISM near the Sun has remained a widely accepted paradigm.

Extended observations of absorption lines toward multiple stars from the Hubble Space Telescope (HST) led to the definition of 15 discrete warm clouds within 15 pc of the Sun (Redfield & Linsky 2008). The ensemble of nearby interstellar material has been denoted the complex of local interstellar clouds (CLIC, Slavin & Frisch 2002). As the velocity vectors of the two clouds were refined with new observations, it was noted that the interstellar flow into the solar system was intermediate between the LIC and G cloud as defined by Redfield & Linsky (2008). Some of the clouds show a filamentary nature, suggesting that they may be created through collisions of compact clouds (Redfield & Linsky 2008). An alternative model with a single cloud with nonrigid flow filling the space within about 9 pc has been proposed (Gry & Jenkins 2014, 2017), but absorption along additional sight lines (Malamut et al. 2014) suggests that the Redfield & Linsky (2008) model better reproduces the HST observations (Redfield & Linsky 2015).

The interplanetary space in our Solar System is filled with the solar wind originating from the Sun, which expands against the VLISM and forms a protective bubble extending more than 100 au from the Sun called the heliosphere (Parker 1958, 1961). Because neutral atoms are not subject to the electromagnetic field in the heliosphere, they travel through the heliosphere on ballistic trajectories. Interstellar neutral (ISN) atoms from the VLISM penetrate the heliosphere, allowing for direct measurements of interstellar material by space experiments (Wallis 1975). The VLISM physical conditions derived from these observations provide the means to understand the Sun's position in the structure of interstellar clouds.

Helium atoms are critical for determining the VLISM properties (Möbius et al. 2004) because of the high universal abundance of helium compared to heavier interstellar species and the high first ionization potential of helium. Consequently, most of the pristine atoms from the VLISM remain neutral on their decades-long journey through the heliosphere (Bzowski et al. 2013; Bzowski & Kubiak 2020). In contrast, hydrogen atoms with lower first ionization potentials become ionized more quickly. Therefore, most hydrogen atoms do not survive transport through the heliosphere to 1 au and instead become ionized and swept out with the solar wind as pickup ions (Möbius et al. 1985; Schwadron 1998).

In this study, we combine the recent results on the VLISM flow obtained from direct sampling of the ISN helium and $He^+$ pickup ion measurements (Section 2). The observations are compared with the flow in the LIC and G cloud, accounting for possible turbulent flow in the immediate neighborhood of the Sun (Section 3). Furthermore, we propose a new concept of mixing interstellar clouds formed by the interaction of the LIC and G cloud to explain the local flow (Section 4). Finally, we discuss how the mixed region explains other observations and summarize our findings (Section 5).

## 2. Interstellar Neutral Flow in the Heliosphere

The flow and temperature of the pristine VLISM around the heliosphere are obtained here from a combination of recent analyses of ISN helium observations. We use direct sampling results from the Interstellar Boundary Explorer (IBEX) and Ulysses missions. Additionally, we constrain the flow direction using the observations of He$^+$ pickup ions from the STEREO-A spacecraft.

The Ulysses mission included the Interstellar Neutral GAS experiment directly sampling ISN helium atoms (Witte et al. 1992). GAS collected ISN data between 1994 and 2007 while Ulysses orbited the Sun in a highly elliptical orbit (aphelion at 5 au and perihelion at 1.3 au), nearly perpendicular to the ecliptic plane. The GAS data have been analyzed in several studies (Witte 2004; Bzowski et al. 2014; Wood et al. 2015).

The IBEX-Lo instrument (Fuselier et al. 2009) is a low-energy atom detector on the IBEX mission, launched in October 2008 (McComas et al. 2009). Analyses of IBEX observations provide the flow velocity relative to the Sun and temperature of the VLISM (Möbius et al. 2009a, 2009b). Because the IBEX pointing restricts the observation of the ISN atoms to close to the perihelion of their trajectories at 1 au and thus to a rather narrow range in ecliptic longitude, the possible interstellar flow speeds and arrival directions in longitude that can be deduced from these observations are tightly coupled through the Keplerian trajectory equation. Likewise, the arrival latitude and temperature are also coupled to form a tube in a 4D parameter space for the ISN (Bzowski et al. 2012; McComas et al. 2012; Möbius et al. 2012). The uncertainty along this IBEX flow degeneracy tube cast doubt on an absolute helium flow determination. Subsequent analyses of the 5-years IBEX data (Bzowski et al. 2015; Leonard et al. 2015; McComas et al. 2015; Möbius et al. 2015a; Schwadron et al. 2015, 2022; Swaczyna et al. 2015, 2018) showed that the most likely ISN helium flow vector is similar to the one inferred through analysis of the GAS data (Witte 2004; Bzowski et al. 2014; Wood et al. 2015).

The most recent analysis of IBEX ISN helium observations (Swaczyna et al. 2022) uses data from 2009 to 2020 and a comprehensive uncertainty system to find the ISN helium parameters. The presented values account for gravitational attraction beyond 150 au (McComas et al. 2015). The parameter tube is represented by the parameter correlation matrix:

$$\boldsymbol{C}_\mathrm{IBEX} = \begin{pmatrix} 1 & -0.844 & 0.316 & 0.962 \\ -0.844 & 1 & -0.279 & -0.861 \\ 0.316 & -0.279 & 1 & 0.364 \\ 0.962 & -0.861 & 0.364 & 1 \end{pmatrix} \begin{matrix} v_\mathrm{IBEX} \\ \lambda_\mathrm{IBEX} \\ \beta_\mathrm{IBEX} \\ T_\mathrm{IBEX} \end{matrix} \qquad (1)$$

where successive columns and rows correspond to the speed ($v_\mathrm{IBEX}$), longitude ($\lambda_\mathrm{IBEX}$), latitude ($\beta_\mathrm{IBEX}$), and temperature ($T_\mathrm{IBEX}$) of the ISN flow. We account for this correlation in our combined parameters.

A recent analysis of STEREO-A/PLASTIC of He$^+$ pickup ions that are ionized inside 1 au provides a needed independent confirmation of the ISN helium flow longitude derived by IBEX (Möbius et al. 2015b; Taut et al. 2018; Bower et al. 2019) and the means to break the degeneracy of the IBEX parameter tube. Because helium atoms are gravitationally accelerated and focused by the Sun, their flow vectors at 1 au follow a distinct pattern that is symmetric about a line parallel to the ISN helium inflow direction toward the Sun. ISN helium atoms, once ionized, are injected into the solar wind as He$^+$ pickup ions with a velocity equal to the difference between the ISN atom velocity and the solar wind bulk velocity. Consequently, the pickup ion velocity distribution cutoff reaches the highest velocity exactly upwind relative to the VLISM flow and the lowest velocity downwind (Möbius et al. 1999), with a smooth and symmetric transition on both sides of the Sun. The STEREO spacecraft sample this structure with each orbit around the Sun. These pickup ion measurements provide an absolute determination of interstellar helium inflow longitude.

The flow longitude was also obtained from SWICS/ACE observations of increased He$^+$ pickup ion density in the helium focusing cone (Gloeckler et al. 2004), but this technique has larger systematic uncertainties (Möbius et al. 2015b). Similarly, the interstellar flow parameters were also obtained from observations of resonant scattering of solar He I 58.44 nm radiation by the ISN helium (Lallement et al. 2004; Vallerga et al. 2004). However, the uncertainties obtained from these studies are about two times larger than those in the studies included in our combination, and thus their statistical significance would be low.

The results from IBEX, Ulysses, and STEREO are listed in Table 1. The combination of the three results is obtained by the minimization of the following sum (Valassi 2003):

$$\chi^2(v, \lambda, \beta, T) = \frac{(\lambda - \lambda_{\text{STEREO}})^2}{\delta \lambda_{\text{STEREO}}^2} + \frac{(v - v_{\text{Ulysses}})^2}{\delta v_{\text{Ulysses}}^2} + \frac{(\lambda - \lambda_{\text{Ulysses}})^2}{\delta \lambda_{\text{Ulysses}}^2} + \frac{(\beta - \beta_{\text{Ulysses}})^2}{\delta \beta_{\text{Ulysses}}^2} + \frac{(T - T_{\text{Ulysses}})^2}{\delta T_{\text{Ulysses}}^2} +$$
$$+ \left[\frac{v - v_{\text{IBEX}}}{\delta v_{\text{IBEX}}}, \frac{\lambda - \lambda_{\text{IBEX}}}{\delta \lambda_{\text{IBEX}}}, \frac{\beta - \beta_{\text{IBEX}}}{\delta \beta_{\text{IBEX}}}, \frac{T - T_{\text{IBEX}}}{\delta T_{\text{IBEX}}}\right] \cdot \boldsymbol{C}_{\text{IBEX}}^{-1} \cdot \left[\frac{v - v_{\text{IBEX}}}{\delta v_{\text{IBEX}}}, \frac{\lambda - \lambda_{\text{IBEX}}}{\delta \lambda_{\text{IBEX}}}, \frac{\beta - \beta_{\text{IBEX}}}{\delta \beta_{\text{IBEX}}}, \frac{T - T_{\text{IBEX}}}{\delta T_{\text{IBEX}}}\right], \quad (2)$$

where $v$, $\lambda$, $\beta$, and $T$ represent speeds, ecliptic longitudes, ecliptic latitudes, and temperatures of the ISN helium obtained in these studies, respectively, while subscripts indicate the mission from which the respective parameters are taken. The analyses of ISN helium observations model fluxes at each spacecraft as a function of these parameters accounting for attraction by solar gravity and ionization losses inside the heliosphere. The best-fit parameters from IBEX are found using $\chi^2$ minimization between the modeled flux integrated over the instrument response and the observed count rates. The analyses (Bzowski et al. 2015; Swaczyna et al. 2022) subtract the contribution from the secondary population created outside the heliopause (Kubiak et al. 2014, 2016; Bzowski et al. 2017, 2019). While the analyses of the Ulysses (Wood et al. 2015) and STEREO (Taut et al. 2018) data did not account for the secondary population directly, they depend on the symmetry of the gravitationally deflected ISN helium distribution, which is little affected by the secondary population (Wood et al. 2017). However, all those studies neglected the elastic collisions that contribute to the filtration beyond the heliopause (Swaczyna et al. 2021). Consequently, the found parameters represent the properties of the primary population at the heliopause, modified with respect to those of the pristine ISN gas by elastic collisions in the outer heliosheath.

Because the parameters derived from IBEX observation are tightly correlated (McComas et al. 2012; Schwadron et al. 2022), as described above, we take into account the covariance matrix. The analyses of Ulysses/GAS observations also provide all considered parameters, but the covariance matrix was not derived (Bzowski et al. 2014; Wood et al. 2015). Nevertheless, the correlations between parameters are much smaller thanks to observations on different parts of the Ulysses orbit leading to the crossing of the correlation tubes (Bzowski et al. 2014). The minimization of the sum given in Equation (2) provides the combined parameters as shown in Figure 1 and listed in Table 1. While the ellipses representing the 1σ uncertainty regions are not always overlapping for the IBEX and Ulysses parameters, the discrepancy is caused by a high correlation between parameters. This is confirmed by the obtained minimum $\chi^2 = 6.69$, which is close to the number of degrees of freedom equaling 5 (9 known parameter values from the considered experiments minus 4 sought parameters), indicating mutual consistency of the observations from these experiments. The statistical probability that the minimum $\chi^2$ is larger than 6.69 is 24.5%. The temperature in the combined result is higher than each of the contributing temperatures. This effect results from the tight parameter tube deduced from IBEX data. The covariance matrix of the combined result calculated from the inverse matrix of the second derivatives of this sum is

$$\boldsymbol{C}_{\text{comb}} = \begin{pmatrix} & v & \lambda & \beta & v \\ 1 & -0.577 & 0.163 & 0.886 \\ -0.577 & 1 & -0.108 & -0.614 \\ 0.163 & -0.108 & 1 & 0.245 \\ 0.886 & -0.614 & 0.245 & 1 \end{pmatrix} \begin{matrix} v \\ \lambda \\ \beta \\ T \end{matrix}. \quad (3)$$

The combined parameters represent the primary ISN helium population, which is already filtered by charge exchange and elastic collisions outside the heliopause. Swaczyna et al. (2021) estimated that the elastic collisions slow down and heat the primary population by $0.45 \pm 0.10$ km s$^{-1}$ and $1100 \pm 300$ K. We account for this small change in the combined ISN parameter results to have a better representation of the pristine VLISM conditions, which we compare with the flows of the interstellar clouds.

Table 1. Flow velocity and temperatures of VLISM and nearby clouds.

| Flow | Reference Frame | Mission | Speed (km s$^{-1}$) | Ecliptic Long.$^a$ (°) | Ecliptic Lat.$^a$ (°) | Galactic Long. (°) | Galactic Lat. (°) | Temperature (K) | Ref. |
|---|---|---|---|---|---|---|---|---|---|
| VLISM | Sun | Ulysses | 26.08±0.21 | 75.54±0.19 | –5.44±0.24 | 183.92±0.23 | –15.26±0.21 | 7260±270 | 1$^b$ |
| VLISM | Sun | IBEX | 25.46±0.21 | 75.69±0.23 | –5.14±0.08 | 183.72±0.13 | –14.98±0.20 | 7250±140 | 2$^b$ |
| VLISM | Sun | STEREO | … | 75.41±0.34 | … | … | … | … | 3$^b$ |
| VLISM | Sun | Combined | 25.71±0.12 | 75.48±0.12 | –5.14±0.07 | 183.62±0.09 | –15.16±0.11 | 7400±80 | 4 |
| Pristine VLISM | Sun | Combined | 26.16±0.16 | 75.48±0.12 | –5.14±0.07 | 183.62±0.09 | –15.16±0.11 | 6300±310 | 4 |
| LIC | Sun | HST | 23.84±0.90 | 78.5±3.2 | –7.2±3.3 | 187.0±3.4 | –13.5±3.3 | 7500±1300 | 5$^c$ |
| G Cloud | Sun | HST | 29.6±1.1 | 71.1±3.0 | –8.5±2.3 | 184.5±1.9 | –20.6±3.6 | 5500±400 | 5$^c$ |
| Pristine VLISM | LIC | … | 2.8±1.1 | 48±22 | 13±23 | 151±29 | –27±26 | … | 4 |
| Pristine VLISM | G Cloud | … | 4.4±1.6 | 220±16 | 28±14 | 13±22 | 55±19 | … | 4 |
| G Cloud | LIC | … | 6.7±1.9 | 43±12 | –12±14 | 172±20 | –46±16 | … | 4 |

**Notes.** $^a$ J2000.0 ecliptic coordinates, $^b$ Source flow direction in ecliptic coordinates, $^c$ Source flow direction in Galactic coordinates.
**References.** (1) Wood et al. (2015), (2) Swaczyna et al. (2022), (3) Taut et al. (2018), (4) *this work*, (5) Redfield & Linsky (2008).

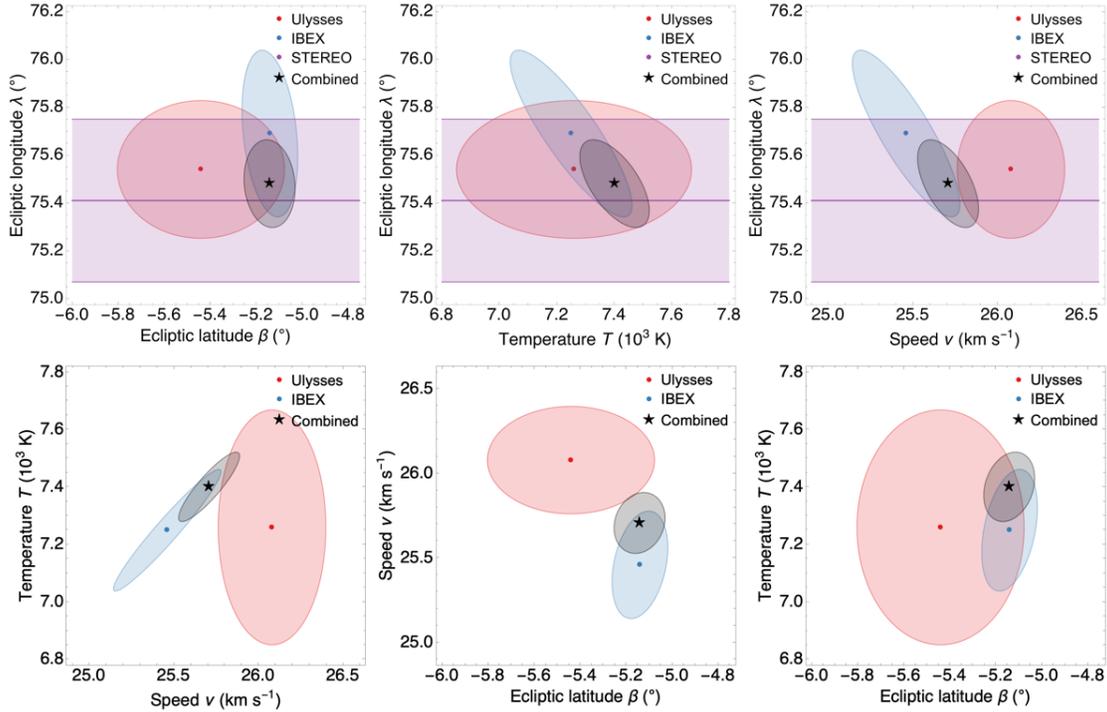

**Figure 1.** Comparison of flow parameter pairs from considered experiments and with the combined result. The ellipses represent uncertainty range and correlations between parameters. The STEREO result constrains only the flow ecliptic longitude. Thus, the top panels show the accepted range as bands, and the bottom panels do not show any STEREO results.

## 3. Inconsistency with the LIC and G Cloud Flows

The flow and temperature of the LIC and G cloud relative to the Sun obtained from the analysis of absorption lines in spectra of nearby stars (Redfield & Linsky 2008) are compared in Table 1 with the combined flow and temperature in the pristine VLISM near the Sun obtained in Section 2. The pristine VLISM flow direction has been transformed into Galactic coordinates to facilitate the comparison. The pristine VLISM flow is significantly different from the flows of the individual interstellar clouds. The table also shows the relative velocity of these clouds in the reference frame of the pristine VLISM and the relative velocity between them.

The LIC also features turbulent intra-cloud motion with a typical magnitude $u_{turb} = 1.62 \pm 0.75$ km s$^{-1}$ (Redfield & Linsky 2008; Linsky et al. 2019), as obtained from a turbulent component of the absorption line broadening. The observed difference between the pristine VLISM and LIC flow speeds is larger than this mean value, and assuming a Gaussian distribution, the probability that the turbulent fluctuation is larger than or equal to $\Delta u_0$ is

$$P(\Delta u \geq \Delta u_0) = \frac{1}{2} - \frac{1}{2} \mathrm{erf} \frac{\Delta u_0 - u_{turb}}{\sigma_{u_{turb}}}. \quad (4)$$

For the observed difference $\Delta u_0 = 2.8$ km s$^{-1}$, this probability is $P(\Delta u \geq 2.8 \text{ km s}^{-1}) = 6\%$. Moreover, if the observed difference in the speed of the pristine VLSIM and LIC is due to a turbulent fluctuation added to the LIC velocity vector, the direction of this turbulent component should be random. The probability that a randomly selected direction on the sphere is at an angle $\alpha$ smaller than $\alpha_0$ is given by the proportion of the solid angle defined by half cone angle $\alpha_0$ to the full solid angle:

$$P(\alpha < \alpha_0) = \frac{2\pi(1 - \cos \alpha_0)}{4\pi}. \quad (5)$$

The actual angle between the pristine VLISM velocity in the LIC frame and the G cloud motion in the LIC frame is $\alpha_0 = 26° \pm 20°$, which gives $P(\alpha < 26°) = 5\%$. The cumulative probability distribution of random fluctuations as a function of the turbulent speed and orientation together is shown in the bottom left panel of Figure 2. The combined probability that a randomly selected turbulent flow vector in the LIC frame differs less from the direction of the G–LIC relative motion than observed and that its speed is larger than observed is just 0.3%, i.e., it is very unlikely that a random turbulent fluctuation can explain the observed VLISM velocity vector in the LIC frame.

## 4. Mixing Interstellar Clouds

Because the flow parameters of the pristine VLISM are approximately mid-way between the LIC and G cloud (Figure 2, right panels), we consider whether a mixture of the LIC and G cloud could explain the flow properties of the VLISM. The mixed region is called by us a mixed interstellar cloud medium (MICM). Let $M_{LIC}$ and $M_G$ denote the contributing masses to the material in the MICM that originate from the LIC and G cloud, respectively. The MICM mass is thus $M_{MICM} = M_{LIC} + M_G$. Conservation of momentum gives $M_{MICM} \boldsymbol{u}_{MICM} = M_{LIC} \boldsymbol{u}_{LIC} + M_G \boldsymbol{u}_G$, where $\boldsymbol{u}_{MICM}$, $\boldsymbol{u}_{LIC}$, and $\boldsymbol{u}_G$ are the bulk velocity vectors of the MICM, LIC, and G cloud, respectively. Using a mixing parameter defined as $\xi = M_{LIC}/M_{MICM}$, this equation can be expressed as $\boldsymbol{u}_{MICM} = \xi \boldsymbol{u}_{LIC} + (1-\xi) \boldsymbol{u}_G$. Therefore, the MICM flow velocity is a linear combination of the flow velocities of these two clouds, with $\xi = 1$ and 0 corresponding to the LIC and G cloud, respectively. Note that this result obtained from the conservation of momentum does not depend on details of the interaction of the material from these two clouds.

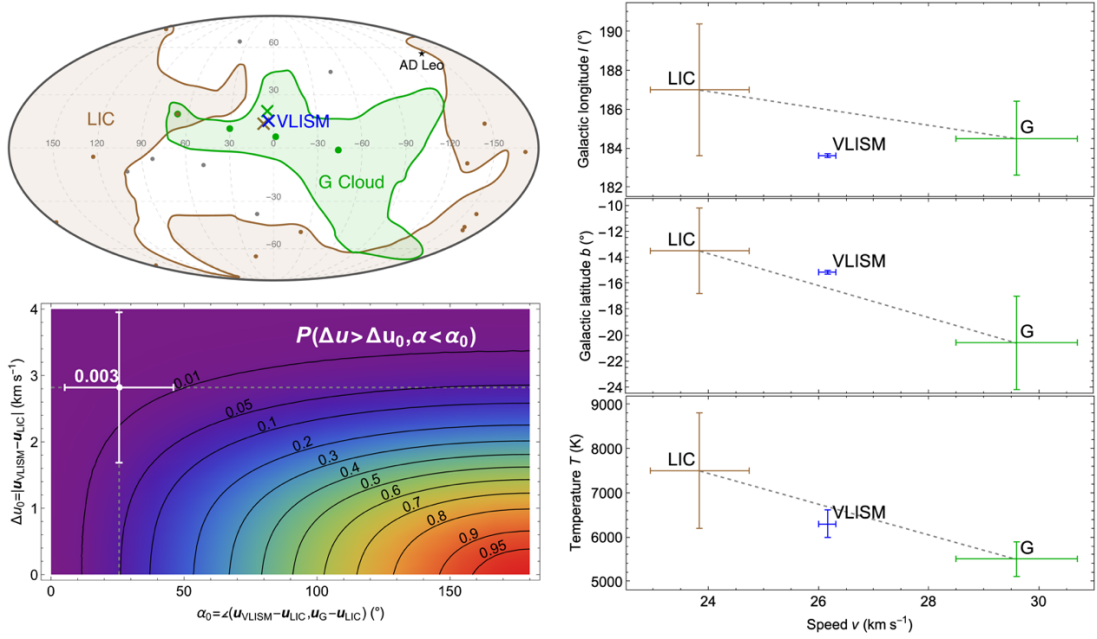

**Figure 2.** *Top left panel:* Spatial extent of the LIC (brown) and G cloud (green) projected on the sky in Galactic coordinates from Redfield & Linsky (2008). The stars within 10 pc from the Sun with identified absorption from nearby interstellar clouds are shown as dots (brown – LIC, green – G cloud, grey – other clouds), the × symbols indicate the inflow directions of the LIC, G cloud, and pristine VLISM. *Bottom left panel:* Color-coded cumulative distribution function of random bulk flow fluctuations inside the LIC (see text). Speed fluctuation is shown on the ordinate and angle away from the G – LIC relative flow vector on the abscissa. The observed speed and angle of the VLISM are marked as a white point with error bars. *Right panels:* Comparison of the flow properties of the LIC and G cloud with the pristine VLISM organized by their speed in the solar frame (top to bottom): Galactic longitude, Galactic latitude, and temperature.

To find the best-fit mixing parameter consistent with the pristine VLISM observations obtained from the space experiments, we minimize the following expression:

$$\chi^2(\xi) = \left[ \boldsymbol{u}_{\text{VLISM}} - \underbrace{(\xi \boldsymbol{u}_{\text{LIC}} + (1-\xi)\boldsymbol{u}_{\text{G}})}_{\boldsymbol{u}_{\text{MICM}}} \right] \cdot (\boldsymbol{\Sigma}_{\text{VLISM}} + \xi\boldsymbol{\Sigma}_{\text{LIC}} + (1-\xi)\boldsymbol{\Sigma}_{\text{G}})^{-1} \cdot$$

$$\cdot \left[ \boldsymbol{u}_{\text{VLISM}} - \underbrace{(\xi \boldsymbol{u}_{\text{LIC}} + (1-\xi)\boldsymbol{u}_{\text{G}})}_{\boldsymbol{u}_{\text{MICM}}} \right], \qquad (6)$$

where $\boldsymbol{\Sigma}_{\text{VLISM}}$, $\boldsymbol{\Sigma}_{\text{LIC}}$, and $\boldsymbol{\Sigma}_{\text{G}}$ denote the covariance matrices of the Cartesian components of the velocity vectors of the pristine VLISM, LIC, and G cloud, respectively. We calculate the Cartesian components of the bulk flows from the parameters provided in Table 1, and the covariance matrices are propagated from the respective uncertainties. Figure 3 presents the expression given in Equation (6) as a function of the mixing parameter. The best-fit mixing parameter is $\xi = 0.62 \pm 0.12$. We repeat the minimization using only the flow vector derived from the IBEX observations to check whether the lack of the Ulysses correlations impacts our findings. The best-fit mixing parameter, in this case, is $\xi = 0.66 \pm 0.13$. Therefore, the best-fit mixing parameter is not strongly affected. We use the value obtained from the combined flow vector for further analysis.

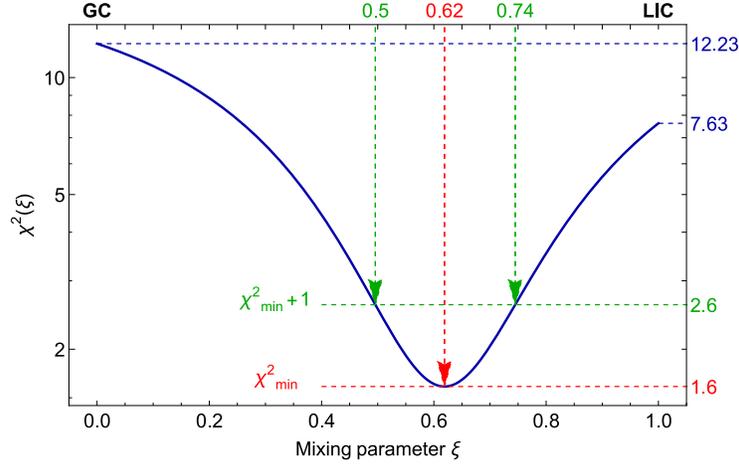

**Figure 3.** Goodness of fit for the mixture model of the LIC and G cloud. The blue line shows the $\chi^2$ value as a function of the mixing parameter value. The best fit is indicated by red dashed lines and the 1σ confidence interval by the green lines.

In addition to the value given by the expression from Equation (6), we also use the Akaike information criterion (AIC) (Akaike 1974) as an additional measure of the goodness of these hypotheses. This criterion accounts for differences in the number of fit parameters ($k$) for the considered hypotheses, and the preferred model has the lowest value of AIC $= 2k + \chi^2$. In our case, the consistencies with the LIC and G cloud are nonparametric hypotheses ($k = 0$), and the MICM has one parameter ($k = 1$). The AIC for the best-fit MICM is 3.6, i.e., much less than AIC for the consistency with the LIC (7.63) and G cloud (12.23) flows. Therefore, we conclude that the MICM is a strongly statistically preferred model consistent with the pristine VLISM flow.

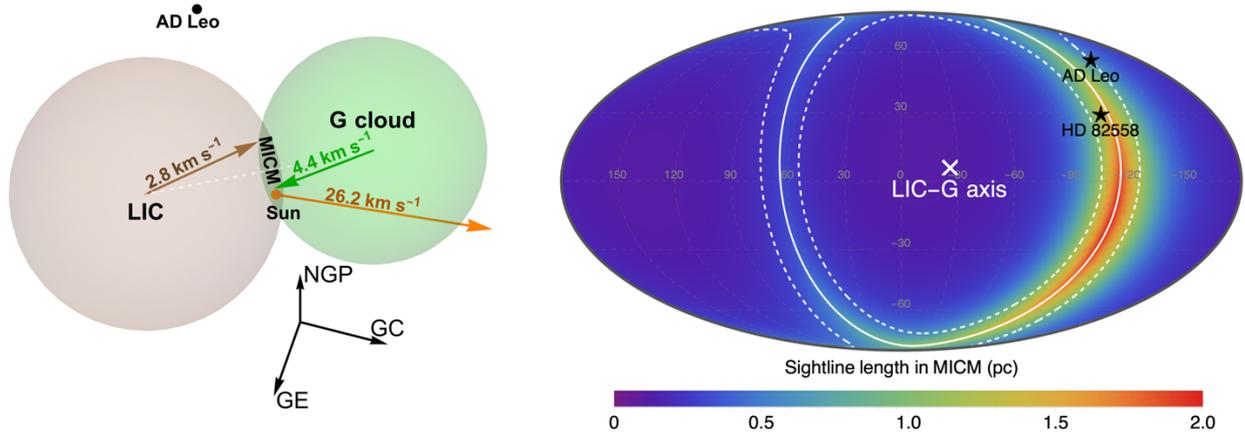

**Figure 4.** *Left panel:* Projection of the illustrative model of the colliding LIC and G cloud forming the MICM from a direction perpendicular to the axis between the clouds. The clouds are represented as spheres with a colliding region called MICM. The orientation relative to the north Galactic pole (NGP), Galactic center (GC), and Galactic east (GE) is shown. Arrows show the relative motion of the clouds in the pristine VLISM frame. The length of the arrow representing the Sun's velocity is reduced 10 times for illustration purposes. The line of sight toward AD Leonis traverses a significant portion of the MICM. *Right panel:* Map in Galactic coordinates of the MICM depth as seen from the Sun, compared with the positions of AD Leonis and HD 82558. The most extended sight lines through the MICM are along the great circle (solid white line) defined by the direction along the line connecting the centers of the LIC and G cloud (white ×).

The mixture model may describe a hypothetical transition region between the LIC and G cloud (Redfield & Linsky 2008) and a collision region between these clouds. The latter possibility is supported by the motion of the G cloud, which overtakes the LIC. Nonetheless, none of the clouds from the Refield & Linsky (2008) model have properties matching the pristine VLISM.

## 5. Discussion and Summary

The density of ISN hydrogen, which was recently found to be $0.195 \pm 0.033$ cm$^{-3}$ in the VLISM near the Sun from observations of hydrogen pickup ions by the SWAP instrument on New Horizons (Swaczyna et al. 2020), additionally supports the MICM. The resulting value is approximately two times higher than the upper limit on the average density of ISN hydrogen ~0.1 cm$^{-3}$ in the local interstellar medium obtained from absorption toward nearby stars located inside the LIC and G cloud (Redfield & Linsky 2008; Linsky et al. 2022).

The limit on the density of ISN hydrogen is obtained by dividing the column density derived from absorption lines by the distance to nearby stars located within the LIC or G cloud. Only two known stars with absorption lines indicate average ISN hydrogen density exceeding 0.13 cm$^{-3}$: AD Leonis and HD 82558. These stars are 5.0 pc (AD Leonis) and 18.3 pc (HD 82558) from the Sun, and the absorption lines indicate column densities of $2.95 \times 10^{18}$ cm$^{-2}$ and $1.122 \times 10^{19}$ cm$^{-2}$ (Redfield & Linsky 2008; Wood et al. 2005), meaning that the average densities of ISN hydrogen along these sightlines are 0.19 and 0.20 cm$^{-3}$. Therefore, these sight lines are filled with a much denser medium than typical sight lines pointing toward the LIC or G cloud. Noteworthy is that AD Leonis is also the closest star for which the absorption was not associated with any cloud near the Sun, while the absorption in the HD 82558 spectrum indicates a component consistent with both the LIC and G cloud (Redfield & Linsky 2008). Furthermore, the positions of these stars are almost perpendicular to the axis connecting the LIC and G cloud centers.

An illustrative model of the colliding LIC and G cloud (Appendix A) is presented in Figure 4. The MICM created between these two clouds has a lenticular shape, and the expected sightline length through the MICM strongly varies with direction in the sky. A substantial depth of the MICM material is expected only within a very narrow angular region, which includes the directions to AD Leonis and HD 82558. Because the actual morphologies of the LIC and G cloud are more complex (Redfield & Linsky 2008), the extinction from the MICM along these directions may exceed the predictions from our model.

The temperature of the pristine VLISM appears to be almost consistent with the LIC and G cloud temperatures owing to the significant uncertainty ranges of these temperatures (Table 1). Therefore, it is not currently possible to constrain the MICM conditions based on the VLISM temperature. Possible LIC – G cloud interaction scenarios predict combined temperatures ranging from $6700 \pm 800$ K to $7300 \pm 900$ K (Appendix B), which are all consistent with the pristine VLISM deduced from the observations of ISN helium atoms.

In addition, a recent study of the IBEX observations revealed that a bi-Maxwellian distribution function with a parallel temperature of $12700 \pm 2960$ K and a perpendicular temperature of $7590 \pm 960$ K to an axis pointing ($170° \pm 7°, –29° \pm 8°$) in Galactic coordinates may better fit the data than a symmetric Maxwellian (Wood et al. 2019). While the authors associated this asymmetry with the interaction with the heliosphere, the angle between this axis and the relative motion of the LIC – G cloud ($172° \pm 20°, –46° \pm 16°$) is $17° \pm 18°$, i.e., slightly smaller than the angle of $19° \pm 7°$ that this axis makes with the pristine VLISM flow direction near the Sun ($183.62° \pm 0.09°, –15.16° \pm 0.11°$), see Table 1. Because mean free paths of helium atoms in the VLISM are a few thousands au (Kubiak et al. 2014), this asymmetry may be explained by ISN helium atoms from the LIC and G cloud penetrating the MICM, which has some probability of reaching the heliosphere. This observation shows that the Sun within the MICM must also be near the main bodies of the LIC and G cloud, consistent with the absorption-line observations.

In summary, the recent observations of ISN helium atoms and He$^+$ pickup ions show that the pristine VLISM near the Sun is a mixture of interstellar materials from the LIC and G cloud. Our conclusion is supported by several observed phenomena, which are consequences of the interaction region between these interstellar clouds. We conclude that the interstellar medium through which the Sun currently travels is not the LIC, and thus the terms LIC and VLISM should not be equated with each other. Instead, we show that the Sun resides in the MICM formed in the interaction of the two nearest interstellar clouds. Consequently, the view of rigid discrete interstellar clouds is not sufficient for characterizing the complexity of the interstellar material in the immediate Galactic environment of the Sun. Because the LIC and G cloud are near the heliosphere at distances that are insufficient to thermalize the ISN helium population fully, the components of the interstellar medium are likely not in equilibrium. Therefore, further studies of the VLISM with a combination of telescopic and in situ observations are needed to find the MICM structure.

*Acknowledgments*: We thank all those who made IBEX, STEREO, Ulysses, and HST possible. P.S. and F.R. are supported by the National Aeronautics and Space Administration (NASA) Outer Heliosphere Guest Investigators Program (grant 80NSSC20K0781). P.S., N.A.S., E.M., P.C.F., D.J.M., F.R., and R.M.W. acknowledge the IBEX mission as a part of the NASA Explorer Program (grant 80NSSC20K0719). E.M. is supported by NASA grant 80NSSC18K1212. S.R. and J.L.L. acknowledge support from the NASA Outer Heliosphere Guest Investigator Program (grant 80NSSC20K0785). R.M.W. acknowledges support from the NASA STEREO TRANSITION grant and grant 80NSSC19K0914. M.B. is supported by grant 2019/35/B/ST9/01241 from the Polish National Science Centre (NCN). B.E.W. acknowledges support from NASA and the Space Telescope Science Institute.

*Data availability*: The IBEX data are available from https://ibex.princeton.edu/RawDataReleases. The STEREO data are available from https://stereo-ssc.nascom.nasa.gov/ins_data.shtml. The Ulysses/GAS data are available from the Ulysses Final Archive http://ufa.esac.esa.int/ufa/. The HST data on the interstellar cloud properties are available in Redfield & Linsky (2008).

## Appendix A. Illustrative Model of the MICM

To illustrate the morphology of the MICM created in the collision of the LIC and G cloud, we model these two clouds as spheres with centers located along the directions of the central coordinates of their positions in the sky, as listed in Table 18 of Refield & Linsky (2008). The LIC and G cloud centers are located along the directions (170°, –10°) and (315°, 0°) in Galactic coordinates, respectively. We assume that the spheres representing these clouds have radii of 3 pc (LIC) and 2.5 pc (G cloud). These numbers are approximately half of the maximum dimensions of these clouds predicted for the ISN hydrogen density of 0.1 cm$^{-3}$ (Lallement et al. 1995; Redfield & Linsky 2000; Linsky et al. 2019).

The Sun is likely in the proximity of the LIC and G cloud because otherwise some absorption from the MICM itself would be visible. Therefore, we assume that the Sun is at 2.9 and 2.4 pc from the LIC and G cloud centers, respectively, i.e., 0.1 pc from the edge of the model spheres. This model defines the interaction region between the LIC and G cloud, which has a lenticular shape elongated perpendicularly to the axis connecting the centers of these clouds, pointing toward (–26°, 6°) in Galactic coordinates.

The details of the illustrative model are uncertain because the positions of the LIC and G cloud centers are at best known with a precision of ~10°. More complex models of the LIC have been developed (Lallement et al. 1995; Redfield & Linsky 2000; Linsky et al. 2019), but these models explicitly precluded the interaction region with the G cloud. Therefore, they cannot be directly implemented in this study.

## Appendix B. Temperature of the MICM

To estimate the temperature of the MICM, we use two limits to describe the character of the LIC – G cloud interaction. These limits are not supposed to provide a detailed physical model of this interaction but are supposed to estimate the possible range for the mixed MICM temperatures. A follow-up study will consider a specific scenario where a compression region between the G cloud and the LIC leads to a temperature and velocity gradient between the two media.

In the first limit, the mixture of local interstellar material is considered to occur without shocks or sudden (irreversible) changes on the kinetic scale. The mixing, therefore, occurs without significantly changing the entropy of the system (isentropic case). In this scenario, the mixed medium is not compressed, and the combined material from each cloud fills a total volume equal to the sum of the contributing volumes from the LIC and G cloud: $V = V_{LIC} + V_G$. The total number of particles in the mixture is also given by the sum of the numbers of particles in each cloud: $N = N_{LIC} + N_G$. Because the elemental composition of both clouds should be similar, we can express the particle number in each cloud using the mixing parameter: $\xi = N_{LIC}/N$, and $1 - \xi = N_G/N$.

The entropy $S$ of an ideal gas is given by the Sackur–Tetrode equation:

$$S = Nk \left\{ \ln\left[\frac{V}{N}\left(\frac{4\pi m U}{3Nh^2}\right)^{3/2}\right] + \frac{5}{2} \right\}, \tag{7}$$

where $U = 3NkT/2$ is the ideal gas internal energy, $h$ is the Planck constant, and $k$ is the Boltzmann constant. Conserving total entropy implies that the total entropy of the mixture is given by the sum of the entropies $S_{LIC}$ and $S_G$ of the individual volumes. With the previous definitions of the densities and the mixing parameter $\xi$ it follows that the temperature of the mixture $T_{MICM}$ is related to the temperatures $T_{LIC}$ and $T_G$ of the individual gases through the following relation:

$$T_{MICM} = \left(\frac{n}{n_{LIC}^{\xi} n_G^{1-\xi}}\right)^{2/3} T_{LIC}^{\xi} T_G^{1-\xi}. \tag{8}$$

Adopting $n_{LIC} \approx n_G$, we find a simplified equation for the temperature

$$T_{MICM} = T_{LIC}^{\xi} T_G^{1-\xi}. \tag{9}$$

The second limit assumes a perfectly inelastic collision of the LIC and G cloud, where the entire loss of kinetic energy by the material from each cloud is dissipated as heat. For the mixing of the two clouds with originally different speeds, the work performed in the process of the mixing can be expressed by the kinetic energy of the two cloud volumes as they approach the final location of the mixed cloud in the reference frame of that cloud, or

$$\Delta W = \Delta W_{LIC} + \Delta W_G = \frac{\xi M_{MICM}}{2}(\boldsymbol{u}_{LIC} - \boldsymbol{u}_{MICM})^2 + \frac{(1-\xi)M_{MICM}}{2}(\boldsymbol{u}_G - \boldsymbol{u}_{MICM})^2, \tag{10}$$

where $\Delta W_{LIC}$ and $\Delta W_G$ are the directional energies of each cloud portion that the clouds shed in the mixing process to end up at the common velocity $\boldsymbol{u}_{MICM}$.

Assuming the helium-to-hydrogen number ratio of $X_{He/H} = n_{He}/n_H = 0.1$ (Frisch et al. 2011), the mean mass of particles in each cloud is

$$\mu = \frac{m_H + X_{He/H} m_{He}}{1 + X_{He/H}} = 1.28 \text{ u}, \tag{11}$$

where u is the atomic mass unit. Therefore, the numbers of particles can be expressed as $N_{MICM} = M_{MICM}/\mu$, $N_{LIC} = \xi M_{MICM}/\mu$, and $N_G = (1-\xi)M_{MICM}/\mu$. The final heat content of the mixed medium is thus

$$\frac{3}{2} N_{MICM} k T_{MICM} = \frac{3}{2} N_{LIC} k T_{LIC} + \frac{3}{2} N_G k T_G + \Delta W. \tag{12}$$

Using Equation (10), the temperature of the MICM is

$$T_{MICM} = \xi T_{LIC} + (1-\xi) T_G + \frac{\mu}{3k}[\xi(\boldsymbol{u}_{LIC} - \boldsymbol{u}_{MICM})^2 + (1-\xi)(\boldsymbol{u}_G - \boldsymbol{u}_{MICM})^2]. \tag{13}$$

Following the expression for the MICM velocity $\boldsymbol{u}_{MICM} = \xi \boldsymbol{u}_{LIC} + (1-\xi)\boldsymbol{u}_G$, this equation can be rewritten as

$$T_{MICM} = \xi T_{LIC} + (1-\xi) T_G + \frac{\mu \xi(1-\xi)}{3k}(\boldsymbol{u}_{LIC} - \boldsymbol{u}_G)^2. \tag{14}$$

The temperature calculated under these two limiting cases (isentropic and perfectly inelastic collision) is presented in Figure A1. The locally observed temperature of the pristine VLISM is consistent with both limits owing to the large uncertainties of the LIC and G cloud temperatures. In our analysis, we use the LIC temperature estimate from Redfield & Linsky (2008) of $7500 \pm 1300$ K (see Table 1). Recently, Linsky et al. (2022) showed that analysis with additional sight lines through the LIC provides a lower temperature of $6500 \pm 500$ K. This new value does not change the consistency of the mixed medium temperature with the pristine VLISM temperature obtained from ISN helium observations.

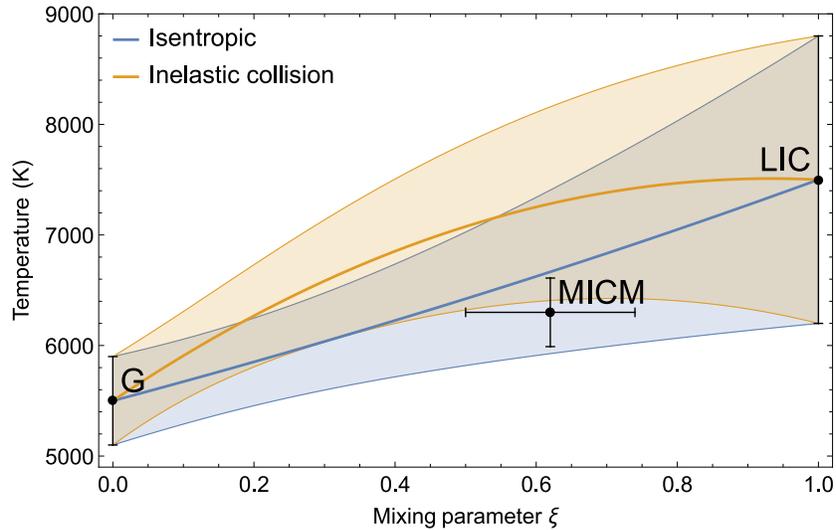

**Figure A1.** Temperature of the MICM obtained from two limiting cases for the mixing of the LIC and G cloud. The bands represent the 1σ uncertainty region. The MICM temperature obtained from local observations is compared to the mixing parameter obtained in this study.